\documentclass[]{aa}  
\usepackage{graphicx,natbib,ulem}
\usepackage{txfonts}
\usepackage{color}

\begin{document}
  \titlerunning{The mass of SS 433}
\authorrunning{ Bowler}
   \title{ The mass of SS 433: a conflict resolved?}

   \subtitle{}

   \author{
          M.\ G.\ Bowler \
          }

  \offprints{M.G.Bowler \\  \email{m.bowler1@physics.ox.ac.uk}}
   \institute{University of Oxford, Department of Physics, Keble Road,
              Oxford, OX1 3RH, UK}
   \date{Received; accepted}

 
  \abstract
   {The Galactic microquasar SS\,433 is very luminous and launches oppositely-directed jets of cool hydrogen at a quarter of the speed of light. Observations of emission lines from the circumbinary disk imply a system mass exceeding 40 $ M_\odot$, with the compact object exceeding 16 $ M_\odot$. The most recent attempts to establish a mass via observation of absorption lines in the spectrum of the companion imply a system mass of less than 20 $ M_\odot$ and a compact object of perhaps 4 $ M_\odot$. }
  {To examine these conflicting data and present a possible resolution of this conflict.}
   {Interpretation of data through the application of simple trigonometry to the configuration of the SS 433 system.} 
  {The absorption spectra which, attributed to the atmosphere of the companion, yield an orbital speed of $\sim$ 60 km s$^{-1}$ could well be attributable to absorption of light from the companion in material of the circumbinary disk. Then the absorption spectra predict an orbital speed for the circumbinary disk material of $\sim$ 240 km s$^{-1}$, in agreement with the emission line data.}
{If continuum light from the companion is absorbed in passage through the circumbinary disk material rather than in the atmosphere of the companion itself, the periodic Doppler shifts in the absorption spectra are entirely consistent with observations of the circumbinary disk and a system mass exceeding $\sim$ 40 $M_\odot$. The  consistency is striking and the implication is that the compact object is a rather massive stellar black hole.}

   \keywords{Stars: individual: SS\,433 - stars: binaries: close - black holes}

   \maketitle
%

\section{Introduction}

The microquasar SS 433 is remarkable for its luminosity and its jets of cool hydrogen ejected almost continuously with a mean speed of about 0.26 $c$. The system is binary and while there is evidence that  the compact object orbits the centre of mass of the binary at $\sim$ 175 km s$^{-1}$ the companion has proved elusive. (The system is reviewed in Fabrika 2004). It is important to establish whether the compact object is a neutron star (unlikely), a black hole of a few solar masses or a much more massive stellar black hole. Observations of emission spectra, interpreted as originating in a circumbinary disk (Blundell, Bowler \& Schmidtobreick 2008, Bowler 2010b), imply a massive black hole; the most recent observations of absorption spectra (Hillwig \& Giess 2008, Kubota et al 2010), when attributed to absorption in the atmosphere of the companion, imply a low mass black hole and the authors are wary of dismissing the possibility of a neutron star. These delicate absorption spectra in the blue show Doppler shifts consistent in phase with an origin in the companion and an orbital speed of $\sim$ 60 km s$^{-1}$. The inferred mass for the compact object is $\sim$ 4 $M_\odot$, for the companion $\sim$ 12 $M_\odot$ and for the system $\sim$ 16 $M_\odot$. My recent analyses of the data on the circumbinary disk (Bowler 2010b) yield an orbital speed for the disk material of $\sim$ 250 km s$^{-1}$ and hence a mass for the system in excess of 40 $M_\odot$. These two measures are not consistent. It can be argued on the one hand that attempts over the years to determine the orbital speed of the companion have yielded little consistency (see the summary in section 6.1 of Kubota et al 2010) and so the absorption spectra can be dismissed as coming from something else; on the other hand that maybe the data interpreted as revealing the circumbinary disk do not in fact yield a reliable estimate of the system mass because the system is full of gas streams. Neither argument is satisfactory in the absence of additional evidence or alternative explanations. In this note I point out that there is a case to be made for the absorption spectra of Hillwig \& Giess (2008) and Kubota et al (2010)
originating as the continuum from the companion is absorbed in circumbinary disk material; the phasing is correct and the orbital speed of the disk then predicted to be $\sim$ 240 km s$^{-1}$, just as obtained from the wholly independent H$\alpha$ and He I emission spectra (Bowler 2010b). This suggestion, if correct, would resolve the conflict, in favour of the companion being a rather massive stellar black hole, as originally argued in Blundell, Bowler \& Schmidtobreick (2008).

\section{The relevant data}

In the work of Hillwig \& Giess (2008) (see also Hillwig et al 2004) and Kubota et al (2010) weak absorption lines appearing the blue part of the spectrum of SS 433 at precessional phase $\sim$ 0, where the accretion disk is most open to the observer, were cross correlated with  spectra of mid A type stars. The results were Doppler shifts from shortly before orbital phase 0 (when the companion eclipses the environs of the compact object) over about a quarter of an orbital period. The extracted recessional velocities have a sinusoidal variation with (semi-)amplitude of $\sim$ 60 km s$^{-1}$ and phase appropriate to an origin in the companion. Furthermore, the visibility of these absorption lines is greatest near eclipse and grows feeble near extreme elongation. These data were interpreted as yielding the orbital velocity of the companion about the binary centre of mass, the assumption being that continuum light from the photosphere of the companion is absorbed in the atmosphere of the companion itself. The systemic speed of these absorption lines is $\sim$ 70 km s$^{-1}$; the widths were interpreted in terms of the companion rotating at $\sim$ 80 km s$^{-1}$ (Hillwig et al 2004).

The stationary emission spectra of SS 433 in both H$\alpha$ and He I are displayed in raw form in Fig.2 of Schmidtobreick \& Blundell (2006) and in terms of the Doppler shifts of fitted Gaussian components in Blundell, Bowler \& Schmidtobreick (2008) and Bowler (2010b). The H$\alpha$ line displays two horns which scarcely change their spacing over more than two orbital periods (but their relative intensity oscillates slightly with the period of the binary orbit). The He I lines correspond to much the same two horns, but with greater variation of intensity. Both H$\alpha$ and He I are consistent with material orbiting in a circumbinary ring and stimulated by some hot spot rotating with the compact object and its accretion disk. The primary evidence for a circumbinary ring is the extreme stability of the H$\alpha$ lines. This stability is particularly striking in Fig.1 of Blundell, Bowler \& Schmidtobreick (2008), where the red and blue lines from the circumbinary disk are contrasted with the periodic motion of the wind in H$\alpha$, which retains a memory of its origin in the accretion disk.The separation of the well marked red and blue horns in both H$\alpha$ and He I is a little under 400 km s$^{-1}$ and as analysed in Bowler (2010b) the orbital speed of the material is $\sim$ 250 km s$^{-1}$. The systemic speed of the circumbinary ring as revealed in emission is $\sim$ 70 km s$^{-1}$. The discussion in Bowler (2010b)
contains references to the most recent data and a critique of supposedly alternative explanations.

Both the absorption line data and the emission spectra attributed to the circumbinary disk are separately convincing but apparently contradictory. A little trigonometry suggests that the contradiction may be only apparent.

\section{A little trigonometry}

The trigonometry is illustrated in Fig.1, for the particularly simple case of equal mass components. The radius at which the companion orbits is $r$ and the angle $\phi$ is the orbital phase, defined to be zero when the companion eclipses the compact object. The companion presents an orbital radius projected on the sky $r$sin$\phi$ as a function of time; the orbit is almost edge on to our line of sight. Thus the companion moves back and forth in the usual projection of circular motion.

\begin{figure}[htbp]
\begin{center}
{\includegraphics[width=12cm]{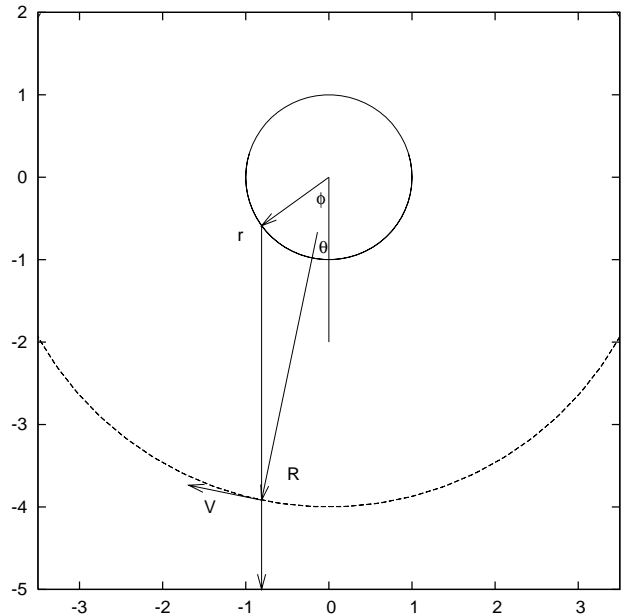}}
\caption{ This diagram shows a plan view of the binary and circumbinary orbits in the SS 433 system. The companion is located at radius $r$ and an azimuthal angle $\phi$, here 0.15 of a period or 54$^{o}$. The companion is viewed through circumbinary material located at $R$, $\theta$ and moving with velocity \textbf{$V$}. The recessional velocity of this material is $V\sin\theta$ and the projections $r\sin\phi$ and $R\sin\theta$ are equal, for all values of $\phi$. The radius of the companion orbit has been taken as $A/2$ and of the circumbinary ring $2A$, where $A$ is the separation of the two binary components.}
\label{fig:beta}
\end{center}
\end{figure}

 The line of sight intersects absorbing material further out in the circumbinary disk, of radius $R$, and that point of intersection moves back and forth across the sky in the same way. When $\phi$ is zero, the companion is viewed (in the centre of mass frame of the binary) through material moving transverse to the line of sight. More generally, for phase $\phi$ the companion is viewed through material with recessional velocity $V_r$ given by

\begin{equation}
V_r=\frac{r}{R}V{\sin\phi}                                                        
\end{equation}

where $V$ is the speed with which the circumbinary material orbits ( $\sim$ 250 km s$^{-1}$) at a radius $R$ =$fA$, $A$ being the separation of the members of the binary. The binary radius $r$ is $\sim$ $A/2$, for mass ratios implied by the orbital speed of the circumbinary ring and $f$ is about $2$. Thus $V_r$ oscillates sinusoidally with an amplitude $\sim$$V$/$4$ so that, if continuum light from the companion is absorbed in the circumbinary disk, the emission line data from the circumbinary disk predict that the absorption lines oscillate with an amplitude of $\sim$ 60 km s$^{-1}$ and the phase of the companion, exactly as observed.

Essentially the same trigonometric argument can be applied to the contribution of the companion size to the width of the absorption lines. The effect matches that for absorption in a companion rotating with speed $aV/R$, where $a$ is the radius of the companion. The radius $a$ is not well determined; if the companion fills its Roche lobe then $a\sim0.4A$, but one sequence of observations suggests it did not then exceed $0.3A$ (Bowler 2010a). Thus the contribution to the line width caused by the size of the companion seen through the circumbinary material matches that for absorption in a companion rotating at  50 km s$^{-1}$ at most; more likely $\sim$ 30 km s$^{-1}$. From observations close to full eclipse Hillwig et al (2004) extracted a rotational velocity for the companion of 80 km s$^{-1}$.

If the argument is inverted, the observed amplitude of the absorption lines, 60 km s$^{-1}$, can be used to predict the orbital speed of the absorbing circumbinary material, as a function of the ratio of the mass of the compact object to that of the companion, $q$.  For a fixed value of the orbital velocity of the compact object ( I took 175 km s$^{-1}$) $q$ and $f$ cannot be chosen independently; thus choosing $q$ determines both $f$ and $V$. The results for $V$ are insensitive to the mass ratio, ranging from a value of 235 km s$^{-1}$ ($f$= 1.69) for $q$ of 0.75 to 250 km s$^{-1}$ ($f$= 2.2) for $q$ as large as 1.1. These values are in agreement with the speed of the circumbinary disk as extracted from the H$\alpha$ and He I stationary emission lines.

Thus absorption in the circumbinary disk can mimic in some detail the expected signature of absorption in the atmosphere of the companion, for an orbital speed of 60 km s$^{-1}$. It is necessary that the right material is present at the right time and in the right condition (also the case for a companion atmosphere) and further that there is no signal necessarily predicted yet not observed.

\section{Discussion - successes}

The observations interpreted as implying an orbital speed for the companion of 60 km s$^{-1}$ were all made for a precession phase within $\sim$0.05 of phase zero. In this configuration the plane of the accretion disk is tilted maximally toward the observer and for orbital phase near 0 the companion appears above the disk and clear of outflowing material in its plane (Hillwig et al 2004). Scattered observations of absorption lines in the blue were presented in Barnes et al (2006); their data were taken with the disk much more edge on and differ in two important respects from those of Hillwig et al (2004), Hillwig \& Gies (2008) and Kubota et al (2010). The scattered data do show approximately sinusoidal variation with orbital phase, but with maximum redshift near orbital phase 0 and the whole pattern shifted to the blue by $\sim$ 100 km s$^{-1}$. These features are attributed to the absorption lines being formed in an equatorial outflow from the accretion disk (Barnes et al 2006, Hillwig \& Gies 2008), for which there is independent evidence (see Fabrika 2004). When the accretion disk is maximally tilted toward the observer, such an outflow is least likely to interfere with observation of absorption lines in the atmosphere of the companion (for orbital phase near 0) and equally is least likely to interfere with observation of absorption in the circumbinary disk.

Hillwig \& Gies (2008) set out three criteria that must be met if their absorption line pattern is to be associated directly with the companion. These criteria may be necessary but they are not sufficient, since the first is met by the completely different origin proposed in this note and the other two may be:

1) The observation of absorption lines in the spectrum of SS 433 oscillating sinusoidally with a period of 13 days and phased to the companion naturally invited interpretation in terms of absorption of continuum light from the photosphere in the atmosphere of the companion. My simple calculations above have shown that if light from the companion is rather being absorbed in material of the circumbinary disk, the lines will oscillate with a period of 13 days, phased to the companion and the numbers work out in remarkable agreement. The systemic radial speeds also agree. Thus the first criterion is met.

2) A number of arguments against these absorption line spectra having a shell like origin are set out in Hillwig et al (2004). It is far from obvious that they apply to the particular case of an orbiting circumbinary
shell. Hillwig et al (2004) remark that a shell like spectrum would likely be seen through a region 
shadowed by the companion from the high flux from the inner accretion disk. This condition is most realised at orbital phase zero and as the phase advances toward extreme elongation irradiation of the relevant region of the circumbinary disk is increasing. Eclipse of the intense source of radiation near the compact object by the companion would relatively enhance absorption lines formed from the companion in the circumbinary disk, just as for formation in the atmosphere of the companion. Thus absorption by the circumbinary disk may not be inconsistent with the variation of absorption amplitude with phase (Fig.3 of Hillwig \& Gies 2008) and the second criterion is likely met.

3) The line broadening due to absorption of light emitted across the diameter of the companion in slightly differently moving parts of the orbiting circumbinary material matches that expected for a synchronously rotating giant companion if absorption takes place in the atmosphere of that giant,   for a rotational speed not exceeding 50 km s$^{-1}$. There could be a problem matching the line widths, because Hillwig et al (2004) inferred a rotational speed of 80 km s$^{-1}$. This potential mismatch might be alleviated by a modest  range of outflow velocities of ring material, the ring being fed by excretion through the L2 point, but this suggestion is highly speculative.

\begin{figure}[htbp]
\begin{center}
\includegraphics[width=12cm]{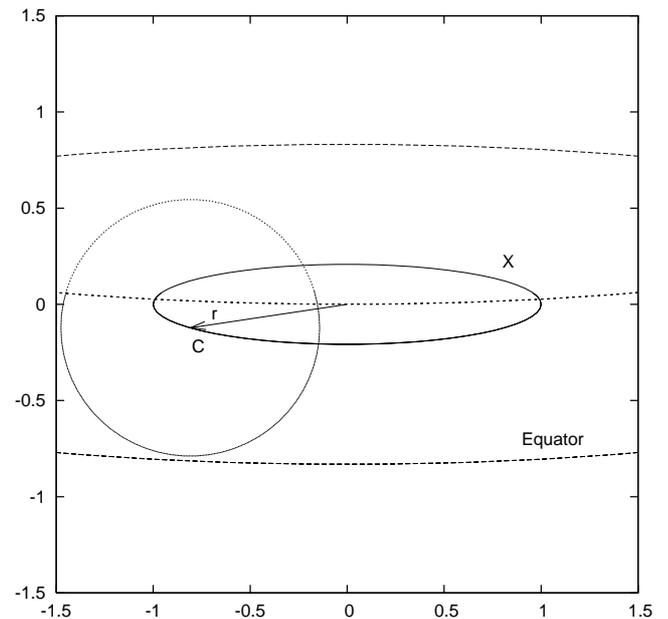}
\caption{This cartoon depicts the extended SS 433 system, looking down at an angle of 12$^{o}$ to the plane of the binary orbit, as observed from Earth. As in Fig.1, a mass ratio of 1 has been assumed and the radius of the binary orbit here seen in projection is $A/2$. The centre of mass of the companion, here denoted by C, is located at \textbf{r} and the compact object X at \textbf{-r} on the opposite side of the orbit. The equator of the circumbinary ring or the core of a torus, of radius $2A$, is shown on both the near and far sides of the system. The shallow curve passing through the origin denotes the latitude of material orbiting $0.4A$ above the orbital plane. The circle illustrates the companion for a radius of $0.3A$; the limb of the companion is peeking over a cloud deck in the circumbinary ring at latitude 12$^{o}$. The companion is otherwise viewed through the circumbinary material.}
\label{fig:delta}
\end{center}
\end{figure}

\section{Discussion - questions}

There remain the questions of whether the right material can be in the right place at the right time and also whether the possibility of absorption lines being formed in the circumbinary disk (or ring or torus) necessarily predicts anything not observed. To clarify the geometry I show in Fig.2 a cartoon of an observer's eye view of the SS 433 system. The line of sight is inclined $12^o$ to the plane of the binary orbit, indicated by the solid ellipse. In this figure and throughout the remainder of this section C denotes the companion and X the compact object. Conditions are as for Fig.1, a mass ratio of unity and orbital phase 0.15 of the orbital period. Sections of the ellipse corresponding to the circumbinary radius $R$ are also shown, marked with the legend "Equator". The shallow curve passing through the origin denotes latitude $12^o$ above the equator of the circumbinary disk. Circumbinary material must reach approximately this latitude for absorption of light from the continuum of the companion to be of significance - a circumbinary disk confined to the equatorial plane would not intercept light from the companion illustrated. This sets out the condition for material being in the right place. It also has to be capable of producing absorption lines in the continuum from the companion, at least for orbital phase 0.85 to 0.25. This is a function of the density of the material, the depth and the temperature, which in turn is likely to depend on the degree of irradiation by X, the accretion disk source. Near orbital phase 0 X is furthest away from material through which the companion is viewed; in addition X is eclipsed for roughly 1 day about phase zero and the material through which C is viewed is in shadow for about 1 day before and 1 day after orbital phase 0. By orbital phase 0.15, C is viewed through material freshly exposed to X. There may thus be an additional reason for the relative amplitude of absorption lines from the continuum of the companion to be feeble out of eclipse.

The second question is whether it is possible to see absorption lines formed from the companion by circumbinary material and not see such lines in the continuum from X. Since X and C are never viewed simultaneously through the same patch of material, it is not necessarily impossible and the question becomes one of plausibility. There is no problem during eclipse but it becomes potentially more troublesome as X emerges from eclipse by C. It can be seen from Fig.2 that if the circumbinary material is too thin to absorb above latitude $10^o$ or so, then visible absorption spectra from X might be avoided to a phase of 0.15 or beyond. Nearer to orbital phase 0.25 the material through which X is viewed has been exposed for a long time to radiation from X itself. Overall, it does not seem implausible that episodes of absorption from the companion in the circumbinary material could be observed without accompanying absorption from X, but it would not be surprising if absorption lines from X became visible roughly 120 km s$^{-1}$ to the blue of those from C, at an orbital phase near 0.25. [Just such a split is visible in one of the spectra of Barnes et al (2006), but those spectra are supposed to have been formed in a somewhat poorly defined outflow and are shifted collectively too far to the blue to be associated with the circumbinary ring (unless an arc had been blown outwards at $\sim$100 km s$^{-1}$ shortly before that spectrum was taken).]

The absorption spectra which are phased with the companion and imply an orbital speed of $\sim$60 km s$^{-1}$ were only observed for precession phases where the accretion disk is most open. The same is true of the spectra analysed by Cherepashchuk et al (2005), where the Doppler shifts are also phased with the companion and match a sine curve with period 13 days. The spectral resolution is inferior to the later observations but if absorption was in the atmosphere of the companion an orbital speed of 132 km s$^{-1}$ is implied, with an error of 9 km s$^{-1}$. Yielding a mass for the compact object of 18$M_\odot$, this would not conflict with inferences from the circumbinary disk. There is agreement that not all the absorption features in the spectrum of SS 433 form in the photosphere of the mass donor companion (Barnes et al 2006, Hillwig \& Gies 2008); the environment in the outer reaches of the system is complicated and it might be that there are absorptive regions more closely associated with the companion, perhaps even within its atmosphere. Different absorptive regions evidently dominate at different times. It might be that a serendipitous sequence of spectra could be companion dominated and it might even be that the data of Cherepashchuk et al (2005) constitute such a sequence.

\section{Conclusions}

The evidence, to my mind compelling, for a circumbinary disk provides an alternative explanation for those absorption spectra hitherto attributed to the atmosphere of a companion orbiting at $\sim$60 km s$^{-1}$. The remarkable agreement between the properties of the circumbinary disk inferred from emission spectra (Bowler 2010b) and from attribution of those weak absorption lines in the blue to material in that same disk suggests very strongly that the compact object in SS 433 is indeed a rather massive stellar black hole, perhaps 20 $M_\odot$ or more. On the evidence now available, it is highly probable that SS 433 contains a circumbinary disk about a binary of total mass greater than 40 $M_\odot$.

The weakest conclusion that could reasonably be drawn is that it is premature to assign the weak absorption spectra observed in the blue, near precession phase 0, to the atmosphere of the companion. Those observations do not necessarily imply a comparatively low mass system, nor that the companion has a mid A type spectrum.

\end{document}